\documentclass[preprintnumbers, prd, showpacs, floatfix,  superscriptaddress,nofootinbib,twocolumn,letterpaper]{revtex4-1}
\usepackage{amsmath, amsthm, amssymb}
\usepackage{latexsym}

\newcommand{\rw}{ {\rm w} }

\begin{document}

\title{General relativity as an attractor for scalar-torsion cosmology}
\author{Laur J\"arv}
\email{laur.jarv@ut.ee}
\affiliation{Institute of Physics, University of Tartu, Ravila 14c, Tartu 50411, Estonia}
\author{Alexey Toporensky}
\email{atopor@rambler.ru}
\affiliation{Sternberg Astronomical Institute, Lomonosov Moscow State University, Moscow 119992, Russia;}
\affiliation{Kazan Federal University, Kremlevskaya 18, Kazan 420008, Russia }


\begin{abstract}
We study flat Friedmann-Lema\^{i}tre-Robertson-Walker cosmological models for a scalar field coupled nonminimally to teleparallel gravity with generic coupling and potential functions.
The goal of this paper is to determine the conditions under which cosmological evolution tends to the limit where the variation of the gravitational ``constant'' ceases and the system evolves close to general relativity. 
These conditions can be read off from the approximate analytical solutions describing the process in matter and potential domination eras.
Only those models where the GR limit exists and is an attractor can be considered viable.
We expect the results to hold in the original ``pure tetrad'' formulation as well as in the recently suggested covariant formulation of the teleparallel theory. In the former case the GR attractor simultaneously provides a mechanism how cosmological evolution suppresses the problematic degrees of freedom stemming from the lack of local Lorentz invariance.

\end{abstract}

\maketitle

\section{Introduction}

Theories with a scalar field nonminimally coupled to gravity as represented by curvature tensor $R$ have a long history and multiple motivations, they arise naturally from compactifications of higher dimensions, in endeavors to build a theory that incorporates the Mach principle or is fundamentally scale-free, or by taking into account quantum corrections to a scalar field minimally coupled to gravity.
In recent years such extensions of general relativity (GR) have received a lot of attention by providing viable models for inflation and dark energy (for a review see e.g.\ Refs. \cite{scalar-tensor review}). In the setup where freely falling test particles follow geodesics (the Jordan frame),  nonminimal coupling manifests itself by making the Newtonian gravitational constant dependent on the value of the dynamical scalar field. However, astrophysical and cosmological observations put a strong limit on the variation of the gravitational constant, e.g.\ since the recombination era $\frac{|G_{\mathrm{rec}}-G_{\mathrm{now}}|}{G_{\mathrm{now}}} < 5 \times 10^{-2}$ \cite{G variation}. 

It was noted by Damour and Nordtvedt \cite{DN} that a large class of nonminimally coupled theories possess an attractor mechanism which during the cosmological evolution makes the solutions to converge to a regime, where the scalar field relaxes around a fixed value, gravitational constant stabilizes, and the theory starts to behave rather like general relativity. This mechanism simultaneously ensures that the stringent constraints from  motions in the Solar System are also met. A method to check whether a nonminimally coupled model spontaneously converges to GR, along with finding the approximate solutions in the vicinity of this limit was developed and applied in Refs.\ \cite{STG attractor}, and was subsequently generalized for arbitrary parametrization of the theory \cite{STG attractor general}.

The same issue must also be addressed in theories with scalar field nonminimally coupled to gravity as represented by torsion tensor $T$ in the framework of teleparallel gravity, which is a possible extension of teleparallel equivalent of general relativity (TEGR) \cite{Geng:2011aj}.\footnote{A similar construction  arises in de Sitter--Cartan geometry with a cosmological function \cite{Jennen:2015bxa}.}
TEGR originates from a theory proposed by Einstein back in 1928 \cite{Einstein}. Instead of Levi-Civita connection (implying zero torsion and non-zero curvature) it uses a Weitzenb\"ock connection \cite{Weitzenb23} (with zero curvature and nonzero torsion), while the description of gravity is based on the tetrad components. 
Although conceptually different, the equations of motion and physical predictions of TEGR are identical to GR, so TEGR can be considered a reformulation of GR \cite{Pereira}. 
But the equations of the extensions of TEGR turn out to be different from their GR counterparts, i.e.\ $f(T)$ gravity from $f(R)$ gravity \cite{Linder}, or a  scalar field nonminimally coupled to torsion from a  scalar field nonminimally coupled to curvature \cite{Geng:2011aj}. This fact makes the study of TEGR extensions worthwhile, as we are exploring completely new ground in the domain of theories.

In the present paper we consider a nonminimally coupled scalar field in a gravity theory on a manifold with Weitzenb\"ock connection, also called scalar-torsion gravity \cite{Geng:2011aj}. 
For specific coupling functions and potentials its flat Friedmann-Lema\^itre-Robertson-Walker (FLRW)  cosmology has been scrutinized by phase space analysis \cite{Wei:2011yr,Xu:2012jf,Otalora:2013tba,Skugoreva:2014ena}, while some analytic solutions were found under particular \textit{Ans\"atze} \cite{Gu:2012ww,Sadjadi:2013nb} or by employing the Noether symmetry method \cite{Noether}. 
Other studies discuss parameter fit with cosmological observations \cite{Geng:2011ka}, growth of density perturbations \cite{Geng:2012vn}, energy conditions \cite{Jamil:2012ck}, and the possibility of singularities \cite{Geng:2013uga}. 

We apply the approach of Refs.\ \cite{STG attractor} to examine flat FLRW cosmologies near the (TE)GR limit where $G$ is constant and the system behaves like in general relativity. In comparison with the usual method of phase space variables rescaled by the Hubble parameter $H$, this approach has some advantages. First, it does not require fixing the form of the coupling function $f$ or the potential $V$ but can treat the completely generic case. Second, it focuses upon the most relevant regime from the phenomenological point of view where the variation of the gravitational constant $G$ is nearly absent. Third, it allows to find approximate analytic solutions for the expansion $H$ and scalar field $\phi$, which are sometimes more amenable for comparison with observations than the rescaled variables. This approach will miss out on the attractors corresponding to evolving $\phi$ (and $G$), but on the other hand it registers fixed points that correspond to the regime where the rescaled variables diverge. 

A problem with the actions of straightforward ``pure tetrad'' TEGR extensions is that they are not invariant under local Lorentz transformations of the tetrad \cite{Lorentz violation}, an indication of preferred frames and extra degrees of freedom which potentially lead to superluminal effects and acausality \cite{Lorentz violation problems, Izumi:2013dca}. As we notice, in the GR limit the terms violating local Lorentz invariance also vanish in the general field equations, so in the models where GR is an attractor, the problematic degrees of freedom get dynamically suppressed by the cosmological evolution.
A very recent proposal to recover local Lorentz invariance is to formulate the theory in a covariant way \cite{Krssak:2015oua} by including purely inertial spin connection (still leading to zero curvature), carefully delineating the effects of gravitation and inertia \cite{renormalized teleparallel}. Since the flat FLRW tetrad in Cartesian coordinates is already ``proper'' \cite{Krssak:2015oua}, we can expect the ensuing cosmological equations not to get additional spin connection contributions and our results will still be valid in the putative covariant formulation of scalar-torsion gravity.

The plan of the paper is as follows. In Sec.\ \ref{sec 2} we briefly recall the foundations of scalar-torsion theory and the equations of motion in flat
FLRW cosmology. Next, the GR limit of the cosmological equations of motion is discussed in Sec.\ \ref{sec 3}. Thereafter we consider two particular situations and
study the problem of reaching the GR limit during a cosmological evolution: Sec.\ \ref{sec 4} is devoted to matter dominated case, and Sec.\ \ref{sec 5} to the scalar field potential dominated case. A further class of solutions, existing in the potential domination case and giving de Sitter solutions are considered separately in Sec.\ \ref{sec 6}. Finally, Sec.\ \ref{sec 7} provides a discussion of the results obtained.

\section{Scalar-torsion cosmology}
\label{sec 2}

Scalar-torsion gravity \cite{Geng:2011aj} extends the teleparallel equivalent of general relativity by introducing a scalar field $\Phi$ nonminimally coupled to torsion, in a manner analogous to how scalar-tensor gravity extends general relativity. Assuming no extra coupling between the scalar and regular matter fields, a generic action of scalar-torsion theory is endowed with three arbitrary functions (compare with Refs.\ \cite{Shapiro:1995kt}, and also \cite{Izumi:2013dca})
\begin{equation}
\!\! S=\int
d^{4}x\,e\,\left[F(\Phi) \frac{T}{2\,\kappa^2}+Z(\Phi) \,\partial_{\mu}\phi\,
\partial^{\mu}\phi-V(\Phi)+\mathcal{L}_m\right] \,.
 \label{generalaction}
\end{equation}
%\textbf{Dimensions not nice, better make $\Phi$ dimensionless?}
Here $\kappa^2 = 8 \pi G$ sets the bare gravitational constant, $\mathcal{L}_m$ is the Lagrangian density of the matter fields, $V(\Phi)$ is the scalar field potential, while $F(\Phi)$ couples the scalar field to gravity as described by the torsion scalar $T$. 

The basic geometrical variables of the theory are the tetrad fields %${\mathbf{e}_A(x^\mu)}$, 
which in a coordinate basis can be expressed as $\mathbf{e}_A=e^\mu_A\partial_\mu$.\footnote{Greek indices $\mu, \nu, \ldots$ run over coordinate spacetime, while capital Latin indices $A, B, \ldots$ span the tangent spacetime.} The tetrads at each spacetime point form an orthonormal basis for the tangent space, they are 
related to the metric tensor through \begin{equation}  
\label{metrvier}
g_{\mu\nu}=\eta_{AB}\, e^A_\mu \, e^B_\nu,
\end{equation}
where $\eta_{AB}={\rm diag} (1,-1,-1,-1)$. The invariant volume element is given by $e = \text{det}(e_{\mu}^A) = \sqrt{-g}$.
%Concerning the independent object that defines the parallel transportation, i.e. the connection, we use 
Assuming the  
Weitzenb\"{o}ck connection $\overset{\mathbf{w}}{\Gamma}^\lambda_{\nu\mu}\equiv
e^\lambda_A\:
\partial_\mu e^A_\nu$ \cite{Weitzenb23} yields zero curvature but nonzero torsion. 
The gravitational field is described by the torsion tensor
\begin{equation}
\label{telelag}
{T}^\lambda_{\:\mu\nu}=\overset{\mathbf{w}}{\Gamma}^\lambda_{
\nu\mu}-%
\overset{\mathbf{w}}{\Gamma}^\lambda_{\mu\nu}
=e^\lambda_A\:(\partial_\mu
e^A_\nu-\partial_\nu e^A_\mu) \,,
\end{equation}
which enters the action as a combination of  contractions,
\begin{equation}
\label{torsionscalar}
T\equiv\frac{1}{4}
T^{\rho \mu \nu}
T_{\rho \mu \nu}
+\frac{1}{2}T^{\rho \mu \nu }T_{\nu \mu\rho }
-T_{\rho \mu }^{\ \ \rho }T_{\
\ \ \nu }^{\nu \mu } \,.
\end{equation}

The action (\ref{generalaction}) is invariant under scalar field reparametrization, 
$\phi=\phi(\Phi)$, thus, for convenience we can adopt a parametrization where one of the functions is a constant. Let us redefine the scalar field so that its kinetic term is canonical,\footnote{In the special case $Z(\Phi)=0$ the action (\ref{generalaction}) is equivalent to the action of $F(T)$ gravity \cite{Izumi:2013dca}. We leave this case out of consideration here.} $Z(\phi)=1$, and split $F(\phi)=1+ \kappa^2 f(\phi)$, so that $f(\phi)$ separates out the dynamical part of the gravitational ``constant.'' This gives the action \cite{Otalora:2013tba,Skugoreva:2014ena}
\begin{equation}
\!\! S=\int
d^{4}x\,e\,\left[\frac{T}{2\,\kappa^2} + f(\phi)\,\frac{T}{2} +\frac{1}{2}\,\partial_{\mu}\phi\,
\partial^{\mu}\phi-V(\phi)+\mathcal{L}_m\right] \,.
 \label{totalaction}
\end{equation}

To study the cosmological solutions corresponding to the flat FLRW spacetime
\begin{equation}
ds^2= dt^2-a^2(t)\,\delta_{ij} dx^i dx^j \,,
\end{equation}
we impose the tetrad \textit{Ansatz} 
\begin{equation}
\label{veirbFRW}
e_{\mu}^A={\rm
diag}(1,a(t),a(t),a(t)),
\end{equation}
where $a(t)$ is the scale factor. 
We also assume that matter is given by a perfect fluid with energy density $\rho_m$ and pressure $p_m$, respectively, related by an equation of state parameter $\rw_m \equiv p_m/\rho_m$. 
Substituting the tetrad and matter content into the equations of motion 
gives rise to the modified Friedmann equations \cite{Otalora:2013tba,Skugoreva:2014ena}
\begin{equation}
\label{Fr1}
3H^2=\frac{\kappa^2}{1+\kappa^2 f(\phi)}\left[\frac{{\dot{\phi}}^2}{2}+V(\phi)+\rho_m\right],
\end{equation}
\begin{equation}
\label{Fr2}
2\dot{H}=-\frac{\kappa^2}{1+\kappa^2 f(\phi)} \left[{\dot{\phi}}^2 +2 H f'(\phi)
\dot{\phi}+\rho_m(1+\rw_m)\right],
\end{equation}
where $H=\dot{a}/a$ is the Hubble function, dots ($\dot{}$) denote differentiation with
respect to the cosmological time $t$, and prime (${}'$) marks differentiation with respect to the scalar field $\phi$.
The scalar field equation of motion is given by
\begin{equation}
\label{phieom}
\ddot{\phi}= -3 H\dot{\phi} - 3  H^2 f'(\phi) -V'(\phi).
\end{equation}
Given that the FLRW tetrad (\ref{veirbFRW}) provides a ``proper tetrad'' \cite{Krssak:2015oua}, i.e.\ it does not need the inclusion of nonzero spin connection coefficients to cancel out inertial effects, we can assume the cosmological Eqns. (\ref{Fr1})-(\ref{phieom}) to hold also in the covariant formulation of scalar-torsion gravity.

Inspection of the Friedmann equation (\ref{Fr1}) reveals that if we assume positive definite potential and matter energy density, $V(\phi)\geq 0$ and $\rho_m \geq 0$, there are real solutions only if
\begin{equation}
\label{f constraint}
\kappa^2 f(\phi) > -1 \,.
\end{equation}
Reaching this bound dynamically would trigger a spacetime singularity, as in the examples studied in Ref. \cite{Geng:2013uga}.

\section{General relativity limit}
\label{sec 3}

If the theory allows the scalar field to take a value $\phi_\star$ such that 
\begin{equation}
 f'(\phi_\star)=f'_\star=0 \,, \qquad V'(\phi_\star)=V'_\star=0 \,,  
\label{grlimit}
\end{equation}
then the Eqs.\ (\ref{Fr1})-(\ref{phieom}) admit a solution $\phi = \phi_\star$, $\dot{\phi}=0$, 
where the cosmological dynamics corresponds to that of (the teleparallel equivalent of) general relativity with matter and cosmological constant set by $V(\phi_\star)=V_\star$.
Here the value $ \frac{1}{8 \pi} \frac{\kappa^2}{1+\kappa^2 f_\star}$ plays the role of gravitational constant $G$ in the cosmological context, as well as in the parametrized post-Newtonian approximation for e.g.\ the Solar System \cite{PPN}. 
The scalar field $\phi$ being constant also makes the problematic terms which are not invariant under local Lorentz transformations of the tetrad to vanish in the general field equation (Einstein's equation) \cite{Izumi:2013dca}, while $f'=0$ makes the pathological degrees of freedom to disappear in an explicit example of nonuniqueness of time evolution \cite{Izumi:2013dca}. 

The aim of the present paper is to assess the stability of these solutions, i.e.\ to determine whether the value $\phi_\star$ functions as an attractor or not. Let us follow the approach and notation of Refs.\ \cite{STG attractor} and expand
\begin{equation}
\label{psi H expansion}
\phi(t)=\phi_\star+x(t) \,, \qquad H(t) = H_\star(t)+h(t) \,, \\
\end{equation}
where $\phi_\star$ is the constant value defined by the condition (\ref{grlimit}) and $H_\star(t)$ is the Hubble function corresponding to the cosmological evolution with $\phi_\star$, while $x(t)$ and $h(t)$ are small perturbations.
It follows that the derivatives are
\begin{equation}
\dot{\phi}(t)=\dot{x}(t) \,, \qquad \dot{H}(t) = \dot{H}_\star(t)+\dot{h}(t) \,.
\end{equation}
 
In the following we are going to apply the expansion (\ref{psi H expansion}) to get approximate field equations which can be solved analytically. Towards this end, recalling the condition (\ref{grlimit}) the functions characterizing the theory get the lowest order contributions as 
\begin{eqnarray}
f(\phi(t))&=& f_\star + \frac{1}{2} f''_\star \, x(t)^2 \,,  \\
f'(\phi(t))&=& f''_\star \, x(t) + \frac{1}{2} f'''_\star \, x(t)^2 \,, \\
V(\phi(t))&=& V_\star + \frac{1}{2} V''_\star \, x(t)^2 \,,  \\
V'(\phi(t))&=& V''_\star \, x(t) +  \frac{1}{2} V'''_\star \, x(t)^2 \,,
\label{DV expansion}
\end{eqnarray}
where the subscript ${}_\star$ denotes the value computed at $\phi_\star$. We assume the derivatives in the expansions above do not diverge at $\phi_\star$.

\section{Matter domination case}
\label{sec 4}

Let us first consider the case when matter energy density dominates over the potential, $\rho_m \gg V(\phi)$, so that we can neglect $V(\phi)$ in Eq.~(\ref{Fr1}). We can use Eq.~(\ref{Fr1}) to eliminate $\rho_m$ in Eq.~(\ref{Fr2}), the result is
\begin{equation}
\begin{split}
2 \dot{H} &+ 3(1+\rw_m) H^2 = \\ &
- \frac{\kappa^2}{1+\kappa^2 f(\phi)} \left[ (1-\rw_m) \frac{\dot{\phi}^2}{2} 
+2 H f'(\phi) \dot{\phi} \right] \,.
\end{split}
\end{equation}
To recover approximate dynamics near the general relativity limit, let us substitute the expansions (\ref{psi H expansion})-(\ref{DV expansion}) into the previous equation and keep terms up to second order small,
\begin{equation}
\begin{split}
2 (\dot{H}_\star + & \dot{h}) + 3 (1+\rw_m) (H_\star + h)^2  = \\ &
 - \frac{\kappa^2}{1+\kappa^2 f_\star} \left[ (1-\rw_m) \frac{\dot{x}^2}{2}  
+2 H_\star f''_\star x \dot{x} \right] \,.
\end{split}
\label{H_h_matter}
\end{equation}
To the lowest order this reduces to
\begin{equation}
2 \dot{H}_\star  + 3 (1+\rw_m) H^2_\star = 0
\end{equation}
which is solved by
\begin{equation}
H_\star(t)= \frac{2}{3(1+\rw_m)(t-t_s)}
\label{H_matter}
\end{equation}
where $t_s$ is a constant of integration that we neglect in the following calculations, $t_s=0$.
The solution (\ref{H_matter}) corresponds to the expansion induced by matter with barotropic index $\rw_m$ in general relativity, as expected.

Substituting the solution (\ref{H_matter}) back into Eq.~(\ref{H_h_matter}) and dropping the terms quadratic in $x$ and $\dot{x}$ gives the equation for the first order correction to the Hubble parameter. It turns out to be simply
\begin{equation}
\dot{h} = - \frac{2 h}{t}
\end{equation}
which is solved by
\begin{equation}
h(t) = h_s t^{-2} \,,
\end{equation}
where $h_s$ is a constant of integration. For any matter barotropic index the cosmological expansion gets asymptotically more close to the one expected in general relativity, as the deviation $h(t)$ decays over time.

The scalar field equation (\ref{phieom}) can be treated in a similar manner, we substitute in the expansions (\ref{psi H expansion})-(\ref{DV expansion}) and the solution (\ref{H_matter}) giving the reference expansion. Keeping only the first order small terms proportional to $x, \dot{x}$, yields an equation of the Bessel type,
\begin{equation}
\ddot{x} = - \frac{2}{(1+\rw_m)t} \, \dot{x} - \frac{ 4 f''_\star}{(1+\rw_m)t^2} \, x - V''_\star x \,.
\label{approx_eq_phi_matter}
\end{equation}

Let us first assume the expansion is around the minimum of the potential, $V''_\star> 0$. Then if the order
\begin{equation}
\nu = \sqrt{\frac{1}{4} \left(\frac{1-\rw_m}{1+\rw_m} \right)^2 - \frac{4}{3} \frac{f''_\star}{(1+\rw_m)^2}  } \,
\label{Bessel order}
\end{equation}
is real the solutions of Eq.~(\ref{approx_eq_phi_matter}) are
\begin{equation}
x(t)  = t^{-\frac{1}{2} \left(\frac{1-\rw_m}{1+\rw_m} \right)} \left( c_1 \, J_\nu (\sqrt{V''_\star} t) + c_2 \, Y_\nu ( \sqrt{V''_\star} t) \right) \,,
\label{approx_sol_phi_matter_real}
\end{equation}
given in terms of the Bessel functions of the first and second kind, $J_\nu$ and $Y_\nu$ respectively, while
$c_1$  and $c_2$ are the integration constants. If the order (\ref{Bessel order}) is imaginary the solutions are still real but are given by \cite{Bessel}
\begin{equation}
\begin{split}
x(t) = t^{-\frac{1}{2} \left(\frac{1-\rw_m}{1+\rw_m} \right)}  \, \mathrm{sech} \frac{\pi \nu i}{2} & \Big( c_1 \, \mathfrak{Re} \left[ J_{\nu} (\sqrt{V''_\star} t) \right] \\ &+ c_2 \, \mathfrak{Re} \left[ Y_{\nu} ( \sqrt{V''_\star} t) \right]  \Big) \,.
\end{split}
\label{approx_sol_phi_matter_imaginary}
\end{equation}

The Bessel functions describe oscillations with slowly decreasing amplitude in time, while the prefactor suppresses the oscillations as a power law when $\rw_m>-1$. This becomes even more transparent when we focus upon late times $t \rightarrow \infty$. Then the Bessel functions admit an asymptotic expansion allowing us to write the approximate solutions as \cite{Bessel}
\begin{equation}
\begin{split}
x(t)  = t^{- \frac{1}{1+\rw_m} } \frac{\sqrt{2}}{\sqrt{\pi} \sqrt[4]{V''_\star} } & \Big( c_1 \, \cos \left( \sqrt{V''_\star} t - \frac{\pi \nu}{2}  - \frac{\pi}{4} \right) \\ & 
 + c_2 \, \sin \left( \sqrt{V''_\star} t - \frac{\pi \nu}{2} - \frac{\pi}{4} \right) \Big) \,,
\end{split}
\label{approx_asym_sol_phi_matter_real}
\end{equation}
for real orders, and
\begin{equation}
\begin{split}
x(t)  = t^{- \frac{1}{1+\rw_m} } \frac{\sqrt{2}}{\sqrt{\pi} \sqrt[4]{V''_\star} } & \Big( c_1 \, \cos \left( \sqrt{V''_\star} t - \frac{\pi}{4} \right) \\ & \,
+ c_2 \, \sin \left( \sqrt{V''_\star} t - \frac{\pi}{4} \right) \Big) \,
\end{split}
\label{approx_sol_phi_matter_imaginary}
\end{equation}
for imaginary orders. So we can conclude that if the matter is not phantom and $V''_\star>0$, then the GR limit acts as an attractor for the solutions near it, since deviations from $\phi_\star$ also vanish over time.

In the case of no potential or at least when $V''_\star$ is zero, the  solutions to
(\ref{approx_eq_phi_matter}) are simpler, namely
\begin{equation}
x(t)= t^{-\frac{1}{2} \left(\frac{1-\rw_m}{1+\rw_m} \right)}  \, \left( c_1 \, t^\nu + c_2 \, t^{-\nu} \right) \,, 
\end{equation}
for real orders $\nu$, and
\begin{equation}
x(t) = t^{-\frac{1}{2} \left(\frac{1-\rw_m}{1+\rw_m} \right)}  \left( c_1 \sin(- i \nu \ln t) + c_2 \cos(- i \nu \ln t) \right) \,.
\end{equation}
for imaginary orders $\nu$ (while $-i \nu$ is real). Again, for imaginary orders the solutions of matter with $\rw_m>-1$ converge to the GR limit. For real orders the solution
consists of two power-law modes, which are both decaying if the matter is not phantom and $f''_\star > 0$. Otherwise one of the power law terms grows in time and the perturbation $x(t)$ does not tend to zero. The latter will eventually invalidate the approximation (\ref{psi H expansion}) and the generic result that the perturbation of the Hubble function $h(t)$ converges will not necessarily hold any more. This is exemplified by the analytic solutions in Refs.\ \cite{Gu:2012ww, Geng:2013uga} for quadratic coupling with dust matter and radiation. 

In the special case when $V''_\star$ and $f''_\star$ are both zero, the first order equation  (\ref{approx_eq_phi_matter}) is left with only a friction term. In order to correctly capture the asymptotic dynamics one has to take into account the leading order force term in the expansion, i.e.\ the lowest nonzero derivative of the potential or coupling function at $\phi_\star$.

Finally, if the expansion (\ref{psi H expansion})-(\ref{DV expansion}) is carried out around the maximum of the potential, $V''_\star< 0$, then the solutions of Eq.~(\ref{approx_eq_phi_matter}) are instead given in terms of the modified Bessel functions of the first and second kind, $I_\nu$ and $K_\nu$. For real orders (\ref{Bessel order}) the solutions read \cite{Bessel}
\begin{equation}
x(t)  = t^{-\frac{1}{2} \left(\frac{1-\rw_m}{1+\rw_m} \right)} \left( c_1 \, I_\nu (\sqrt{-V''_\star} t) + c_2 \, K_\nu ( \sqrt{-V''_\star} t) \right) \,,
\label{approx_sol_phi_matter_real}
\end{equation}
while for imaginary orders 
\begin{equation}
\begin{split}
x(t)  = t^{-\frac{1}{2} \left(\frac{1-\rw_m}{1+\rw_m} \right)}  \, \Big( & c_1 \, \mathfrak{Re} \left[ I_{\nu} (\sqrt{-V''_\star} t) \right] \\ &  + c_2 \, \mathfrak{Re} \left[ K_{\nu} ( \sqrt{-V''_\star} t) \right]  \Big) \,.
\end{split}
\label{approx_sol_phi_matter_imaginary}
\end{equation}
The modified Bessel function of the first kind diverges asymptotically as $I_\nu (t) \sim \frac{e^t}{\sqrt{2 \pi t}}$, therefore in the case of $V''_\star< 0$ the GR limit acts as a repeller for all solutions nearby.

To recap, in the matter dominated case ($\rho_m \gg V(\phi)$) and for non phantom equation of state ($\rw_m > -1$) the solutions spontaneously converge towards the GR values $\phi_\star$ and $H_\star$  either if the GR limit corresponds to the minimum of the potential $V''_\star>0$ or for vanishing $V''_\star$ to the minimum of the scalar-torsion coupling, $f''_\star > 0$.

\section{Potential domination case}
\label{sec 5}

Let us now consider the case when the scalar field potential dominates over the matter energy density, $V(\phi) \gg \rho_m$. Then the Friedmann equation (\ref{Fr1}) in the lowest order yields de Sitter background,
\begin{equation}
\label{H de Sitter}
H_\star =  \sqrt{ \frac{\kappa^2 V_\star}{3 (1+\kappa^2 f_\star)} } \,,
\end{equation}
while in the next order we see that the deviation $h(t)=0$.
Expanding the $\phi$ equation (\ref{phieom}) around the general relativity limit (\ref{grlimit}) and substituting  (\ref{H de Sitter}) gives to the first order
\begin{equation}
\label{x pot dom eq}
\ddot{x} = - 3 H_\star \dot{x} - 3 H^2_\star f''_\star x - V''_\star x \,.
\end{equation}
Its solution, given in terms of the constant
\begin{equation}
v_\star= \sqrt{ 1- \frac{4 f''_\star}{3} - \frac{4 (1+ \kappa^2 f_\star) V''_\star}{3 \kappa^2 V_\star} } \,,
\end{equation}
is
\begin{equation}
x(t) = e^{-\frac{3 H_\star t}{2}} \left( c_1 e^{ \frac{3 H_\star v_\star t}{2}}+ c_2 e^{ - \frac{3 H_\star v_\star t}{2}} \right)
\end{equation}
for real $v_\star$, and
\begin{equation}
x(t) = e^{-\frac{3 H_\star t}{2}} \left( c_1 \sin { \frac{3 H_\star i v_\star t}{2}}+ c_2 \cos { \frac{3 H_\star i v_\star t}{2}} \right)
\end{equation}
for imaginary $v_\star$. The solution for the deviations $x(t)$ decays asymptotically to zero if
\begin{equation}
\label{pot dom gen rel condition}
f''_\star + \frac{(1+ \kappa^2 f_\star) V''_\star}{\kappa^2 V_\star} > 0 \,.
\end{equation} 
Therefore, the minimum of the potential, $V''_\star>0$, coinciding with the minimum of the scalar-torsion coupling, $f''_\star>0$, guarantees convergence to genreal relativity, however, Eq.\ (\ref{pot dom gen rel condition}) also allows for the possibility that only one of these functions is at its minimum there. (The constant $\frac{(1+ \kappa^2 f_\star)}{\kappa^2 V_\star}$ is positive, cf.\ Eqs.~(\ref{f constraint}) and (\ref{H de Sitter}).) The fixed point $J$ in Ref.\ \cite{Xu:2012jf} in the case of quadratic coupling and constant potential is a particular example of this class of solutions.

For mathematical completeness let us note that if $\rho_m \equiv 0$ and $V_\star=0$ the reference background is equivalent to Minkowski, $H_\star=0$. The first order expansion of the Friedmann equation (\ref{Fr1}) vanishes identically, while in the second order it gives
\begin{equation}
h(t)^2 = \frac{\kappa^2}{6(1+\kappa^2 f_\star)} \left( \dot{x}^2 + V''_\star x^2 \right) \,.
\end{equation}
The first two orders in the expansion of the equation for the scalar field (\ref{phieom}) read
\begin{equation}
\label{x pot dom V=0}
\ddot{x} = -3 h \dot{x} - V''_\star x - \frac{1}{2} V'''_\star x^2 \,.
\end{equation}
If $V''_\star \neq 0$ the force term proportional to $x$  dominates, while the friction term proportional to $h \dot{x}$ is subdominant. Therefore in the leading approximation we get oscillating solutions for $V''_\star>0$,
\begin{equation}
\label{x sol Minkowski V''>0}
x(t)= c_1 \sin(\sqrt{V''_\star}t) + c_2 \cos(\sqrt{V''_\star}t) \,,
\end{equation}
or exponentially diverging solutions for $V''_\star<0$,
\begin{equation}
\label{x sol Minkowski V''<0}
x(t)= c_1 \, e^{\sqrt{-V''_\star}t} + c_2 \, e^{-\sqrt{-V''_\star}t} \,.
\end{equation}
This result must be taken with caution, however, since the subleading Hubble friction term will also have an effect on the oscillations (\ref{x sol Minkowski V''>0}) in the long run.
Such an effect was first described for a minimally coupled scalar field with quadratic potential, $V=V_0 \phi^2$, in general relativity \cite{starobinsky1978}.
In fact, since the contribution from  nonminimal coupling, $f''_\star$, is of even lower order and does not occur in Eq.\ (\ref{x pot dom V=0}), the situation here is exactly like in Ref.\ \cite{starobinsky1978}, where a WKB approximation was used to establish 
\begin{equation}
x_{t \rightarrow \infty} \sim \frac{\sin \sqrt{2 V_0}t}{\sqrt{2 V_0}t} \,.
\end{equation}
Also note that if $V>0$ around $V_\star$, the Hubble parameter $h$ will not change its sign due to Eq. (\ref{Fr1}). Therefore in an expanding universe ($h>0$) the friction term has the correct sign, the oscillations will gradually decrease as described above and the system relaxes to Minkowski, while for a contracting universe ($h<0$) the friction term has an opposite sign causing instead an amplification of the oscillations, until the approximation scheme breaks down. 
In the case when $V''_\star=0$ also, all first order terms vanish in Eq.\ (\ref{x pot dom V=0}) and further analysis is needed to understand the dynamics.

\section{Balanced solutions in the potential domination case}
\label{sec 6}

It is interesting to notice that in the case when the matter density can be neglected in the equations, there is another possibility to have a static solution for the  scalar field in Eqs.\ (\ref{Fr1}) and (\ref{phieom}), namely if the theory allows for the existence of such $\phi_*$ that satisfies the condition of balance
\begin{equation}
\label{balance condition}
- \kappa^2 V_* f'_* - V'_* (1+\kappa^2 f_*) =0 \,,
\end{equation}
where $V_*=V(\phi_*)$ etc. Then $\dot{\phi_*}=0$ and the system conforms with general relativity while the background is again de Sitter with
\begin{equation}
H_* = \sqrt{\frac{\kappa^2 V_*}{3 (1+\kappa^2 f_*)} } 
= \sqrt{-\frac{V'_*}{3 f'_*}} \,.
\end{equation}
Note that this de Sitter solution is not ``encoded'' into the initial Lagrangian of the theory because this $\phi_*$ in general does not
correspond to a minimal value of either the potential or the coupling function. The cosmological constant associated with this solution is an effective one, appearing due to an interplay between the scalar field potential and the nonminimal coupling.

Expanding Eqs.\ (\ref{Fr1}) and (\ref{phieom}) around this value, $\phi(t) = \phi_* + x(t)$, $H(t)=H_*+h(t)$, and taking into account the condition (\ref{balance condition}) gives for the small deviations
\begin{equation}
h(t)= \frac{\kappa V'_*}{\sqrt{3 V_* (1+\kappa^2 f_*)} } \, x(t)
\end{equation}
and
\begin{equation}
\label{x eq bal case}
\ddot{x} = - 3 H_* \dot{x} - 3 H^2_* f''_* x - V''_* x - \frac{2 f'_* V'_*}{1+\kappa^2 f_*} x \,.
\end{equation}
The latter is solved in terms of the constant
\begin{equation}
v_* = \sqrt{1- \frac{4 f''_*}{3} - \frac{4}{3} \frac{(1+\kappa^2 f_*) V''_*}{\kappa^2 V_*} - \frac{8}{3} \frac{f'_* V'_*}{V_*} }
\end{equation}
by 
\begin{equation}
x(t) = e^{-\frac{3 H_* t}{2} } \left( c_1 e^{\frac{3 H_* v_* t}{2}} + c_2 e^{-\frac{3 H_* v_* t}{2}} \right) 
\end{equation}
for real $v_*$, and by
\begin{equation}
x(t) = e^{-\frac{3 H_* t}{2} } \left( c_1 \sin{\frac{3 H_* i v_* t}{2}} + c_2 \cos{\frac{3 H_* i v_* t}{2}} \right) 
\end{equation}
for imaginary $v_*$.
Hence, if the scalar-torsion theory under consideration admits such $\phi_*$ which satisfies the condition (\ref{balance condition}), then $\phi_*$ acts as an attractor for the nearby solutions and the system converges to general relativity if
\begin{equation}
 f''_* + \frac{(1+\kappa^2 f_*) V''_*}{\kappa^2 V_*} +  \frac{2 f'_* V'_*}{V_*}
> 0 \,. 
\end{equation}
These results generalize the de Sitter fixed points described for power law scalar-torsion couplings and exponential potentials \cite{Wei:2011yr, Xu:2012jf, Otalora:2013tba}, as well as power law potentials \cite{Wei:2011yr,Skugoreva:2014ena}.

Finally, let us note that in the absence of matter it is mathematically possible to solve the condition (\ref{balance condition}) also by $V_*=0$, $V'_*=0$. Then the reference background is Minkowski, $H_*=0$, and the approximate equation (\ref{x eq bal case}) coincides with Eq.\ (\ref{x pot dom V=0}) considered before.

\section{Discussion}
\label{sec 7}

In this paper we considered flat FLRW cosmologies in generic scalar-torsion gravity, with focus upon the limit where the variation of the gravitational constant ceases and the system behaves akin to general relativity.
Comparing with the usual cosmological phase space analysis in terms of the variables rescaled by the Hubble parameter, our method nicely reproduces the fixed points (and their properties) which correspond to a constant value $\phi_\star$ of the scalar field, while those fixed points which correspond to running scalar field (e.g.\ $\phi$ tending to infinity) are not seen in our analysis by construction.
On the other hand, our method is able to find and treat the fixed points where $H_\star=0$ and the rescaled variables diverge.
By employing a Taylor expansion near a fixed point, our method proceeds without assuming a fixed form of the coupling function $f$ and potential $V$, hence our results generalize the earlier studies \cite{Wei:2011yr, Xu:2012jf,Otalora:2013tba,Skugoreva:2014ena}.
Still, as we considered mostly the first order in the expansion, some potentially interesting models where the lower derivatives of the coupling function and potential vanish at the fixed point (like $\phi^4$ at $\phi_\star=0$) were left out of our present inquiry. These can be treated by going to higher orders in the expansion.

Motivated by the cosmological limits on the variation of the gravitational ``constant'' $G$, our goal was to determine the conditions under which a model under investigation is endowed with general relativity as an attractor.
A viable model should have a constant scalar field value $\phi_\star$ which maintains its attractive character though radiation, dust matter, and potential domination eras. 
The respective conditions turn out to be quite natural. The existence of $\phi_\star$ is given by the coinciding extrema of  the coupling and potential functions, 
$f'_\star=V'_\star=0$. The property of attraction to GR for nonphantom matter ($\rw_m>-1$) domination era is ensured if $V''_\star>0$, or if $V''_\star=0$ then at least $f''_\star>0$. For the (positive definite) potential domination era the attraction is provided by $f''_\star + \frac{(1+ \kappa^2 f_\star) V''_\star}{\kappa^2 V_\star}>0$, or if $V_\star=0$ then by $V''_\star>0$. Therefore a sufficient condition for attraction is that both the coupling and potential functions have a minimum there, although in principle it is possible to also achieve attraction when the coupling function has its maximum but the minimum of the potential is ``stronger'' (in the sense of the conditions above).

We may assess the viability of some typical couplings and potentials popular in the literature. For instance an outcome of this analysis is that with exponential potential functions it is not very easy to construct viable models, as these foster a GR limit only in the potential dominated era (the balanced case at $\phi_*$).  
If in the matter dominated era we may neglect all effects from the exponential potential and have a GR regime at $\phi_\star$ due to suitable (e.g.\ power law) coupling, then the observational restrictions on the variability of G mean that $\phi_\star$ and $\phi_*$ must be rather close to each other, which puts a constraint on the allowed values of the model parameters.

From constant or power law coupling $f=\xi \phi^N$ and potential $V=V_0 \phi^n$, however, we can expect GR like behavior near $\phi_\star=0$, for $N, n>2$. There are various combinations for how the GR limit can be an attractor through all the eras. For a vanishing or positive constant potential, a quadratic coupling function with a positive parameter $\xi$ leads to and keeps the system near the GR limit, while a negative $\xi$ makes the GR limit unstable. Nonnegative quadratic potential ($V_0>0$) ensures that the GR limit is an attractor irrespective of the power index and sign of the coupling $f$. So, a small negative $\xi$ in quadratic coupling which seems to be favored by observations \cite{Geng:2011ka} is consistent with GR attractor behavior if the potential has a quadratic term, i.e.\ the scalar field has a mass. Higher powers in the coupling and potential need further examination, since one has to go to higher orders in the Taylor expansion.

As mentioned in the Introduction, adding a scalar field nonminimally coupled to teleparallel gravity in the ``pure tetrad'' formulation makes the action to lose invariance under local Lorentz transformations of the tetrad,
indicating that the tetrad fields incorporate additional and possibly problematic degrees of freedom, extra to the usual metric ones \cite{Lorentz violation, Lorentz violation problems, Izumi:2013dca}.
As anticipated, these extra degrees of freedom
do not manifest themselves at the level of FLRW  background solution \cite{Geng:2011aj}. 
Here we note that in the GR limit, where the scalar field tends to a constant value, the Lorentz invariance violating terms disappear also in the full equations of motion. Therefore viable models where GR is an attractor are automatically equipped with a dynamical mechanism whereby cosmological evolution suppresses the problematic nonmetric degrees of freedom.

In the very recently proposed covariant formulation of TEGR extensions \cite{Krssak:2015oua}, the ``pure tetrad'' approach is augmented by including possibly nonzero spin connection which removes the spurious inertial effects and restores local Lorentz invariance.
Since the flat FLRW tetrad is already ``proper'' \cite{Krssak:2015oua} and the respective cosmological equations do not acquire corrections from including the spin connection, we can expect the results of our analysis to remain valid also in the covariant formulation of scalar-torsion gravity.

In the literature there are several proposals to modify scalar-torsion gravity: making the scalar field tachyonic \cite{tachyonic}, coupling the scalar field nonminimally to $F(T)$ gravity \cite{nonminimal F(T)}, coupling torsion to the scalar kinetic term \cite{Kofinas:2015hla}, introducing a boundary term \cite{Bahamonde:2015hza}, or combinations that make the theory conformally invariant \cite{conformal}. We expect our method and reasoning to also work in these models, although the precise results should depend on the particular theory under consideration.

\begin{acknowledgments}

L.J. was supported by the Estonian Research Council Grant No.\ IUT02-27 and by the European Union through the European Regional Development Fund (Project No.\ 3.2.0101.11-0029). A.T. was supported by RFBR Grant No.\ 14-02-00894, and was partially supported by the Russian Government Program of Competitive Growth of Kazan
Federal University. The authors thank Margus Saal and Emmanuel N. Saridakis for useful discussions, as well as the anonymous referee for pointing out a typo.

\end{acknowledgments}

\end{document}